# Time-reversal symmetry-breaking phenomena in transport study of kagome superconductivity

Shuo-Ying Yang[†], Jia-Xin Yin

(Department of Physics, Southern University of Science and Technology,

Shenzhen 518055, China)

**Abstract**

Recent studies have found that all three materials within the vanadium-based kagome superconductors ($A$V$_3$Sb$_5$, $A$ = K, Cs, Rb) exhibit time-reversal symmetry-breaking behaviors in the superconducting states. Among the three, the Josephson junctions structured Nb/K$_{1-x}$V$_3$Sb$_5$/Nb and RbV$_3$Sb$_5$ show magnetic hysteresis below the superconducting transition temperature. CsV$_3$Sb$_5$ exhibits a zero-field superconducting diode effect, meaning the magnitude of the positive and negative superconducting critical currents are different. We first discuss the similarities and differences among the three above-mentioned experiments. Then, we discuss the possible mechanisms responsible for the unconventional superconducting transport phenomena: such as a chiral superconducting order parameter, and chiral pair density waves arising from the intertwining of the chiral charge order with superconductivity.

† corresponding author email: yangsy@sustech.edu.cn

# 1. Introduction

Symmetry plays a fundamental role in solid state physics[1]. Since symmetry breaking often leads to the emergence of novel states of matter, experimental techniques for detecting symmetry breaking are essential. A well-known example is the Hall effect, which is the transverse voltage generated by a longitudinal electrical current. Most Hall effects result from an external magnetic field or intrinsic magnetic order that breaks time-reversal symmetry (TRS). The Hall effect has stimulated extensive research in topological quantum states of matter, fractional charge quantization, and the application of quantized resistance in the International System of Units. When TRS is preserved, the first-order Hall effect disappears. However, if inversion symmetry (IS) is broken, the system can exhibit a nonlinear Hall effect, characterized by a nonlinear relationship between the Hall voltage and the driving current[2].

Superconductivity has emerged as one of the most active research areas in condensed matter physics in recent years. The scientific community is actively looking for new types of superconducting materials, while simultaneously seeking evidence for unconventional superconductivity. For instance, d-wave pairing is believed to be associated with curprate high-temperature superconductivity. The p-wave spin-triplet superconductors can generate dissipationless spin currents and can also carry topological non-trivial Majorana fermions.

In recent years, a series of novel quantum states of matter driven by topology, electron correlation, and magnetic interaction have been discovered in vanadium-based kagome superconductors ($AV_3Sb_5$, $A$ = K, Cs, Rb)[3]. Studies have found that all three materials within the $AV_3Sb_5$ family exhibit superconductivity. However, the unusual feature of the kagome superconductivity has been elusive[4]. Superconductivity in three-dimensional systems often has both TRS and IS. For the superconducting critical current ($I_c$), having IS means $I_c$ follows $|I_c^+(\Phi)| = |I_c^-(\Phi)|$ (where $I_c^+$ is the $I_c$ at positive current, $I_c^-$ is the $I_c$ at negative current, and $\Phi$ represents the magnetic flux). Having TRS requires $|I_c^+(\Phi)| = |I_c^-(-\Phi)|$. If IS is maintained and TRS is broken, the system exhibits $|I_c^+(\Phi)| \neq |I_c^+(-\Phi)|$, that is, the magnetoresistance in the superconducting state depends on the magnetic field sweep direction, but not on the polarity of the current. If TRS is maintained and IS is broken, the system exhibits $|I_c^+(\Phi)| \neq |I_c^-(\Phi)|$, meaning under an external magnetic field, the magnitude of $I_c$ is related to the polarity of the current. If both IS and TRS are missing, a non-reciprocal superconducting state will appear at zero magnetic field ($|I_c^+(\Phi=0)| \neq |I_c^-(\Phi=0)|$). Therefore, hysteresis in the field and/or current sweep can used to analyze symmetry breaking in the superconducting state.

This view and perspective will focus on three low-temperature quantum transport experiments, aiming at studying the symmetry-breaking superconducting states in $K_{1-x}V_3Sb_5$, $CsV_3Sb_5$, and $RbV_3Sb_5$.

## 2. Unconventional superconducting transport phenomena in vanadium-based kagome superconductors

One early quantum transport indicating possible symmetry-breaking superconductivity was reported by Ali *et al.*, who studied Josephson junctions of $Nb/K_{1-x}V_3Sb_5/Nb$ (x~0.26-0.31)[5]. These sets of samples are not intrinsic superconducting, possibly because they are K-deficient. However, upon putting superconducting Nb electrodes on the exfoliated flakes, the samples exhibit superconductivity below 0.93 K. The authors observed several interesting features in the proximity-induced superconductivity. Firstly, they noted that the magnetoresistance in the superconducting state depends on the magnetic field sweep direction, but is independent of the current polarity. Such an asymmetry suggests the system exhibits IS but breaks TRS. Secondly, the Fraunhofer patterns resulting from the interference of two superconducting wavefunctions in the Josephson junction are highly anisotropic. It shows a typical Fraunhofer-like pattern for an in-plane magnetic field, but an anomalous pattern with a minimum near zero-field for an out-of-plane applied field. This not only implies that the internal magnetic field for $K_{1-x}V_3Sb_5$ is anisotropic, but also suggests the proximity-induced cooper pairs might be spin-triplet, similar to what is observed in $CrO_2$[6]. Lastly, the Fraunhofer pattern also reveals a rapid oscillating period with magnetic field, akin to the double-slit interference phenomenon seen in optical experiments. Those long-lived, fast oscillations could potentially originate from the spatially localized topological states.

In a recent work done by Lin *et al.* at the Westlake University[7], it was discovered that intrinsic superconducting $CsV_3Sb_5$ flakes exhibit a superconducting diode effect at zero magnetic field and a superconducting interference pattern resulting from flux quantization. A notable finding from this experiment is the observation of the superconducting diode effect at zero magnetic field in a single superconducting material. This observation indicates that the material simultaneously breaks both IS and TRS[8]. In addition, the polarity of the superconducting diode effect can be reversed after thermal cycling just above the superconducting transition temperature, possibly due to the modulation of superconducting domains with temperature. The second interesting phenomenon is the double-slit-like interference pattern. Since this pattern originates from a single superconducting sample, its origins are likely different from the Josephson junction in $K_{1-x}V_3Sb_5$. The authors attributed the source of this superconducting interference pattern to the flux quantization inside the superconductor - the Little-Parks effect, which results from the formation of closed-loop supercurrent[9]. In this effect, the kinetic energy of superconducting electrons varies periodically with the increasing magnetic field, ultimately manifesting as a periodic change in the critical temperature as a function of the magnetic field. Similar features have also been reported in the intrinsic Weyl superconductor $MoTe_2$[10]. The combination of the superconducting diode effect and the double-slit superconducting interference pattern suggests the presence of boundary superconducting currents in $CsV_3Sb_5$ with broken TRS and IS.

More recently, Lin *et al.* at the Shenzhen Institute of Quantum Research observed unconventional superconducting behavior in $RbV_3Sb_5$[11]. The unconventionality is evident in two main aspects. Firstly, there is hysteresis behavior in the magnetoresistance within the superconducting state, which appears only under the in-plane magnetic field, indicating the presence of spontaneous TRS breaking in the system. It is worth emphasizing that this hysteresis behavior appears exclusively under an in-plane magnetic field, which contrasts with the hysteresis behavior observed in $Nb/K_{1-x}V_3Sb/Nb_5$ Josephson junctions, where it manifests under out-of-plane magnetic field[5]. In addition, the temperature-magnetic field ($T$-$B$) phase diagram under ithe n-plane magnetic field shows two superconducting domes. These domes exhibit different responses under an out-of-plane magnetic field, suggesting that they may have different origins. Furthermore, the in-plane field can transition one superconducting state into another, exhibiting "reentrant" superconducting behavior. The $T$-$B$ phase diagram also reveals that at temperatures above 600 mK, the superconducting state only emerges under an in-plane magnetic field, similar to what has been observed in $UTe_2$[12] and Bernal bilayer graphene[13]. The authors believe that the in-plane magnetoresistance hysteresis and "reentrant" superconductivity are manifestations of spin-polarized spin-triplet superconductivity.

## 3. Discussion and Outlook

Table 1 summarizes the main findings from the above three transport experiments. Among the three materials studied, $K_{1-x}V_3Sb_5$ exhibits proximity-induced superconductivity, while both $CsV_3Sb_5$ and $RbV_3Sb_5$ are intrinsic superconductors. Both the Josephson junction of $Nb/K_{1-x}V_3Sb_5/Nb$ and the $CsV_3Sb_5$ display superconducting interference patterns due to spatially localized superconducting currents. $CsV_3Sb_5$ is the only material among the three that shows IS breaking, while all three materials exhibit TRS breaking. Hysteresis in magnetoresistance is observed in both $Nb/K_{1-x}V_3Sb_5/Nb$ and $RbV_3Sb_5$, although they occur under different magnetic field directions.

The transport properties of the three $A$V$_3$Sb$_5$ material systems show some similarities, but there are also notable differences. From the example of $K_{1-x}V_3Sb_5$, we have learned that the superconducting properties of vanadium-based kagome superconductors are highly sensitive to the sample's stoichiometry. The superconducting diode measurement of $CsV_3Sb_5$ shows that different electrodes of the same sample have different $I_c$ behavior, and the distribution of superconducting domains varies with different thermal cycles. Whether this randomness is associated with the inhomogeneity of the sample, and to what extent the stoichiometry of the sample affects the different quantum phases and their competition with one another are questions worthy of consideration.

Table 1. Comparison of superconducting transport properties of different vanadium-based kagome superconductors.

|  | Nb/K$_1$-V$_3$Sb$_5$/Nb[5] | CsV$_3$Sb$_5$[7] | RbV$_3$Sb$_5$[11] |
| --- | --- | --- | --- |
| Device type | Josephson junction | Intrinsic superconductor | Intrinsic superconductor |
| Spatially localized supercurrent | √ Topological edge state | √ Little-Parks effect | — |
| Inversion symmetry breaking | — | √ Superconducting diode effect | — |
| Time reversal symmetry breaking | √ out-of-plane hysteresis | √ Superconducting diode effect | √ in-plane hysteresis |

The thickness dependence of these materials also warrants further exploration. Recent studies have shown that exfoliated samples with thickness in the tens of nanometers exhibit different transport characteristics compared to bulk materials. In samples approximately 20 nm or less, the magnetoresistance shows linear behavior below the charge density wave (CDW) transition temperature, and a quadratic dependence on the magnetic field above the CDW transition temperature. However, when the thickness exceeds 20 nm, the low-temperature linear magnetoresistance transitions to a sublinear dependence[14]. The intrinsic superconducting characteristics were already reported in bulk materials back in 2020[15-17]. What causes the non-reciprocal superconducting transport phenomena to occur only in thin films of several tens of nanometers? How does the change in thickness affect symmetry breaking in the superconducting state? These are questions that merit further investigation.

Additionally, the three studies do not reach a unified conclusion regarding the anisotropy of these materials. For instance, Nb/$K_{1-x}V_3Sb_5$/Nb and $CsV_3Sb_5$ show TRS breaking in an out-of-plane magnetic field, while $RbV_3Sb_5$ exhibits TRS breaking in an in-plane magnetic field. The source of this anisotropy remains to be determined. Other than the anisotropy observed in the superconducting state[18,19], in-plane magneto-transport measurements have also detected anisotropic magnetoresistance in the non-superconducting state of $CsV_3Sb_5$[20]. The relationship between the anisotropy of superconductivity and the anisotropy of CDW state and normal system, as well as their distinct origins, requires further clarification.

Furthermore, even though all three materials in the $AV_3Sb_5$ family exhibit TRS-breaking, the exact mechanism responsible for the TRS-breaking in the superconducting state is unclear. A number of theories have proposed possible mechanisms, including chiral flux phases[21], charge bond orders[22], and spin density waves[23], although all these are associated with non-superconducting states. In addition to the aforementioned electrical transport experiments, recent STM measurements have revealed that the superconducting order parameter in $AV_3Sb_5$ all exhibit chiral 2×2 pair density waves. This finding suggests the presence of an unconventional intertwined order. As such, whether the symmetry breaking in the superconducting phase originates from chiral superconducting order parameters, such as d+id[25] or p+ip[11] pairing, or from chiral pair density waves generated by the coupling of chiral CDW and conventional superconducting states, are important questions to be addressed in future studies.

Vanadium-based kagome superconductors exhibit complex and rich ordered states that are highly sensitive to the electronic states near the Fermi level. Given their intricate phase diagram and the competition among various quantum states, it is crucial to identify effective methods to modulate these states - such as chemical or electrical doping, stress, or pressure. Isolating and manipulating a single controlling parameter through these techniques will help in understanding its complex phase diagram. In addition, further investigation is needed to verify and analyze the observed unconventional superconducting transport properties. This includes but is not limited to exploring its thickness dependence, anisotropy, the relationship between CDW and the superconducting state, as well as the origin of IS and TRS states.

To investigate the thickness dependence, one could systematically fabricate devices of varying thicknesses and examine how these variations influence symmetry breaking within the superconducting state. Research into anisotropy would require a thorough characterization of the system's rotational symmetry under different crystallographic orientations, magnetic fields, and quantum states, analyzing the interrelationships among them. Studies on the relationship between CDW and superconducting states could involve using Ta-doped $Cs(V_{1-x}Ta_x)_3Sb_5$. Previous research has shown that Ta-doped $Cs(V_{1-x}Ta_x)_3Sb_5$ exhibits a higher critical temperature than $CsV_3Sb_5$, alongside the disappearance of the CDW phase[26]. Comparing the non-reciprocal transport in $Cs(V_{1-x}Ta_x)_3Sb_5$ and $CsV_3Sb_5$ could shed light on whether CDW is a necessary condition for symmetry breaking. To unveil the intrinsic order parameters of superconductivity, phase-sensitive experiments are needed. This includes measuring the Little-Parks effect in mesoscopic rings to elucidate the charge and spin pairing of the Cooper pairs[27], as well as fabricating corner SQUID devices combined with conventional s-wave superconductors to investigate the symmetry of order parameter[28].

Currently, research into the unconventional superconducting properties of vanadium-based kagome superconductors has only revealed the tip of the iceberg. We look forward to uncovering the underlying mechanism through a more comprehensive chain of evidence.